\documentclass[lettersize,journal]{IEEEtran}
\usepackage{amsfonts}
\usepackage{amsmath}
\usepackage{algpseudocode}
\usepackage{algorithm}
\usepackage{array}
\usepackage[caption=false,font=normalsize,labelfont=sf,textfont=sf]{subfig}
\usepackage{textcomp}
\usepackage{stfloats}
\usepackage{url}
\usepackage{verbatim}
\usepackage{graphicx}
\usepackage{cite}
\usepackage{nomencl}
\usepackage{etoolbox}
\usepackage{setspace}
\usepackage{amsthm}
\usepackage{color}  
\usepackage{afterpage} 

\usepackage[bottom=1.8cm, top=1.8cm,left=1.4cm,right=1.4cm]{geometry}
\makenomenclature
 
\renewcommand\nomgroup[1]{%
  \ifthenelse{\equal{#1}{A}}{\item[\textbf{\textit{Sets and Indices}}]}{%
  \ifthenelse{\equal{#1}{B}}{\item[\textbf{\textit{Parameters}}]}{%
  \ifthenelse{\equal{#1}{C}}{\item[\textbf{\textit{Variables}}]}{ }}}}
 
\nomenclature[A]{$S$, $s$}{Set and index of scenarios.}
\nomenclature[A]{$L$, $ij$}{Set and index of lines.}
\nomenclature[A]{$N$, $i$}{Set and index of buses.}
\nomenclature[A]{$T$, $t$}{Set and index of time.}
\nomenclature[B]{$c_{l}$}{Investment cost of hardening line $l$}

\nomenclature[B]{$c_{b}$}{Unit investment cost of the energy storage.}
\nomenclature[B]{$p_{s}$}{Probabilities of each scenario $s$.}
\nomenclature[B]{$c_{g}$}{Operation cost of generator $g$.}
\nomenclature[B]{$c_{w}$}{Wind curtailment cost of wind farm $w$.}
\nomenclature[B]{$c^{ch/dis}$}{Charging \negthinspace/ \negthinspace Discharging cost of the energy storage.} 
\nomenclature[B]{$c^{N/E}_{LS}$}{Cost of load shedding under normal \negthinspace/ \negthinspace extreme condition.} 
\nomenclature[B]{$C^{L}$}{Line investment budget.} 
\nomenclature[B]{$C^{B}$}{Energy storage investment budget.}
\nomenclature[B]{$\Delta{t}$}{Time interval.}
\nomenclature[B]{$\eta^+/\eta^-$}{Charging \negthinspace / \negthinspace Discharging efficiency of the energy storage.}
\nomenclature[B]{$P^{\rm Max/Min}_{g}$}{Maximum \negthinspace / \negthinspace Minimum output of generator $g$.}
\nomenclature[B]{$P^{\rm Max}_{ij}$}{Maximum capacity of line $ij$.}
\nomenclature[B]{$P_{i,t,s}^{D,N/E}$}{Load on bus $i$ at time $t$ under normal \negthinspace/ \negthinspace extreme condition in scenario $s$.}
\nomenclature[B]{$B_{ij}$}{Susceptance of line $ij$.}
 
\nomenclature[B]{$\theta_{i,t,s}$}{Angle of bus $i$ at time $t$ in scenario $s$.}
\nomenclature[C]{$P^{g,N/E}_{i,t,s}$}{Active power output of the generator at bus $i$ at time $t$ under normal \negthinspace/ \negthinspace extreme condition in scenario $s$.} 
\nomenclature[C]{$Z_{i}$}{Capacity of the energy storage at bus $i$.}

\nomenclature[C]{$P^{dis,N/E}_{i,t,s}$}{Discharging power of the energy storage at bus $i$ at time $t$ under normal/extreme condition in scenario $s$.} 
\nomenclature[C]{$P^{ch,N/E}_{i,t,s}$}{Charging power of the energy storage at bus $i$ at time $t$ under normal \negthinspace/ \negthinspace extreme condition in scenario $s$.} 
\nomenclature[C]{$P^{LS,N/E}_{i,t,s}$}{Operation cost of generator $g$.}
\nomenclature[C]{$P_{ij,t,s}^{N/E}$}{Active power on line $ij$ at time $t$ under normal \negthinspace/ \negthinspace extreme condition in scenario $s$.}
\nomenclature[C]{$P_{i,t,s}^{w,N/E}$}{Forecasted wind outputs on bus $i$ at time $t$ under normal \negthinspace/ \negthinspace extreme condition in scenario $s$.}
\nomenclature[C]{$P_{i,t,s}^{wc,N/E}$}{Wind curtailment on bus $i$ at time $t$ under normal \negthinspace/ \negthinspace extreme condition in scenario $s$.}
\nomenclature[C]{$x_{ij}$}{Binary variables indicating the hardening decision of line $l$. (1 if line $ij$ is hardened; 0 otherwise).}
\nomenclature[C]{$x^{dis,N/E}_{i,t,s}$}{Binary variables indicating the discharging status of the energy storage on bus $i$ at time $t$ under normal \negthinspace/ \negthinspace extreme condition in scenario $s$. (1 if the energy storage is discharging; 0 otherwise).}
\nomenclature[C]{$x^{ch,N/E}_{i,t,s}$}{Binary variables indicating the discharging status of the energy storage on bus $i$ at time $t$ under normal \negthinspace/ \negthinspace extreme condition in scenario $s$. (1 if the energy storage is charging; 0 otherwise).}
\nomenclature[C]{$S_{i,t,s}^{N/E}$}{State of charge of the energy storage on bus $i$ at time $t$ under normal \negthinspace/ \negthinspace extreme condition in scenario $s$.}
\nomenclature[C]{$\mu_{ij,t,s}$}{Binary variables indicating the status of line $ij$ at time $t$ in scenario $s$. (1 if line $ij$ is damaged; 0 otherwise).}
\nomenclature[C]{$\gamma_{w,t,s}$}{Binary variables indicating the status of wind farm $w$ at time $t$ in scenario $s$. (1 if wind farm $w$ is damaged; 0 otherwise).}
\nomenclature[C]{$\alpha_i/\beta_i$}{Load shedding ratio of bus $i$ under normal and emergency scenarios.}
\nomenclature[C]{$\varepsilon_{ij,t,s}^w/o$}{Binary variables indicating the status of line $ij$ at time $t$ in scenario $s$ with \negthinspace/ \negthinspace without line hardening. (1 if line $ij$ is damaged; 0 otherwise)}

\begin{document}

\title{Risk-informed Resilience Planning of Transmission Systems Against Ice Storms}

\author{Chenxi Hu,~\IEEEmembership{Student Member,~IEEE}, Yujia Li,~\IEEEmembership{Member,~IEEE,}, Yunhe Hou,~\IEEEmembership{Senior Member,~IEEE,}\vspace{-1.5em}
\thanks{This work was supported by National Key R\&D Program of China (2023YFA1011301), Joint Research Fund in Smart Grid (Grant No. U1966601) under cooperative agreement between the National Natural Science Foundation of China (NSFC) and State Grid Corporation of China (SGCC), National Natural Science Foundation of China (NSFC) under Grant 52177118.}}



\maketitle

\begin{abstract} 
Ice storms, known for their severity and predictability, necessitate proactive resilience enhancement in power systems. Traditional approaches often overlook the endogenous uncertainties inherent in human decisions and underutilize predictive information like forecast accuracy and preparation time. To bridge these gaps, we proposed a two-stage risk-informed decision-dependent resilience planning (RIDDRP) for transmission systems against ice storms. The model leverages predictive information to optimize resource allocation, considering decision-dependent line failure uncertainties introduced by planning decisions and exogenous ice storm-related uncertainties. We adopt a dual-objective approach to balance economic efficiency and system resilience across both normal and emergent conditions. The first stage of the RDDIP model makes line hardening decisions, as well as the optimal sitting and sizing of energy storage. The second stage evaluates the risk-informed operation costs, considering both pre-event preparation and emergency operations. Case studies demonstrate the model's ability to leverage predictive information, leading to more judicious investment decisions and optimized utilization of dispatchable resources. We also quantified the impact of different properties of predictive information on resilience enhancement. The RIDDRP model provides grid operators and planners valuable insights for making risk-informed infrastructure investments and operational strategy decisions, thereby improving preparedness and response to future extreme weather events. 

\end{abstract} 

\begin{IEEEkeywords}
Resilience enhancement, decision-dependent uncertainty, transmission system planning, stochastic optimization.
\end{IEEEkeywords}
  
\begin{spacing}{0.92} 
\printnomenclature
\end{spacing} 

\section{Introduction} 


Power systems are vulnerable to weather-driven catastrophes, such as hurricanes, ice storms, and floods, resulting in significant economic loss and heightened risks to public safety \cite{7801854}. Among them, ice storms are noteworthy due to their long duration and wide scope. Ice accretion and heavy snow can cause system failure from line overloading, galloping, and icing flashover. 
Additionally, the strong winds combined with ice accumulation can greatly influence power generation and cause major disruptions, especially with the increasing penetration of renewable energy sources \cite{9913670}.
In 2008, the ice disaster in Southern China led to more than \$50 billion in economic losses and the failure of thousands of power lines and towers \cite{8818698}. More recently, Texas was struck by the winter storm Uri in February 2021, resulting in \$130 billion in losses and about 2000 MW wind power went offline\cite{BUSBY2021102106}. There is a growing concern that the frequency and severity of such events could increase as a result of climate change, increasing the exposure of power systems to failures \cite{furman2013economic}. 
This emphasizes the importance of enhancing power system resilience on the ice disaster prevention capabilities.

During ice storms, transmission lines are susceptible to ice accretion which can lead to large-scale blackouts. Effective resilience enhancement measures include the application of anti-icing materials, the use of de-icing technologies, and the development of de-icing operation strategies\cite{9496094}. Apart from operational strategies, this research emphasizes the importance of incorporating resilience considerations as early as the planning stage.
Typical strategies include robust structure design, or relocating facilities to make the system less prone to disruptive events. A key hardening strategy against ice storms is the use of anti-icing coating materials\cite{farzaneh2022techniques}, which enables the transmission line to resist thicker ice. Additionally, strategically relocating facilities, such as bulk energy storage, can not only optimize the system operation under normal conditions but also help provide a continuous electricity supply for critical loads during outages. 
The distributed characteristics of energy storage as well as its load-shifting capability make it naturally suitable for normal operation and resilience enhancement during emergencies \cite{zhang2017optimal}. While those measures can reduce the chance of component failures and the needed restoration efforts, hardening and upgrading the entire system is expensive. Therefore, how to prioritize transmission lines for hardening, and how to allocate the energy storage to balance costs and system resilience have become great challenges that require further research.

Despite numerous resilience enhancement measures, achieving effective resilience planning must be crafted with awareness and understanding of the characteristics of different kinds of system failure. Those differences mainly lie in two aspects \cite{national2017enhancing}: (1) how much predictive information, e.g., event types, start time, duration and trajectory, system operators have to make preparations; (2) how much of the systems remain controllable and operable during contingencies. While resilience enhancement with limited resources has been widely investigated \cite{li2021restoration,qiu2017integrated}, existing research seldom underlines the effect and potential of predictive information on the resilience enhancement of power systems. With advanced forecasting technologies, we are able to predict extreme weather events, including ice storms \cite{degelia2016overview}. 
This predictive information is pivotal for preparatory, preventative, and remedial actions. 
In this context, the term \textit{predictive information} is defined as the information that quantifies insights derived from available information to anticipate future events \cite{bialek1999predictive}. 
The lead time and accuracy of predictive information can influence the operational decision-making process by furnishing valuable insights before events, thereby facilitating preparation and engendering varied response strategies during contingencies.
However, much of the existing research often treated predictive information as deterministic inputs, overlooking the inherent uncertainty of the information itself.
Ma et al \cite{ma2016resilience} proposed a tri-level resilience enhancement strategy by incorporating the coupling of the hardening decisions and the damage uncertainty. Li et al \cite{li2023distributionally} proposed a scenario-wise decision-dependent ambiguity set to model the uncertainty brought by preventive resilient enhancement strategies and the related extreme weather-induced system contingencies. In fact, given the current technological capabilities and anticipated advancements, the accuracy and reliability of information can vary significantly. The availability of information can greatly influence the decision-making process, leading to diverse investment strategies. 
In this regard, a comprehensive risk-informed planning model should not only account for uncertainties stemming from investment decisions but also consider the uncertainties intrinsic to predictive information.

Beyond the previously mentioned hardening measures, the impact of the planning decisions on the operating states of the power system is another usually overlooked yet significant part. Previous resilience planning research mainly focused on emergency scenarios during extreme events, ignoring the preceding risk-informed conditions wherein the events are already anticipated, given certain predictive information. 
Though some research has encompassed operation during various operational conditions, a primary concern is that these methods neglect the transition in operational objectives between normal and emergency conditions.
Hossein et al. \cite{ranjbar2021resiliency} proposed a resilience planning model for transmission systems considering both normal and emergency conditions using distributed energy resources. Lagos et al.\cite{lagos2019identifying} proposed a framework for resilient network investments against earthquakes using an optimization via simulation approach. Though their models acknowledge varying operational states, these conditions are treated independently and the coordination between the normal and emergency states is not considered. 
During contingencies, there is a paradigmatic shift in operational objectives. The priority transitions from economic operation to an emphasis on retaining critical loads, rather than ensuring the adequacy of all loads. This significant change underscores the inadequacy of traditional planning approaches and highlights the necessity for a coordinated strategy that can pre-dispatch available resources. 

Based on the previously discussed research gaps, a two-stage risk-informed decision-dependent resilience planning (RIDDRP) model for transmission systems against ice storms is proposed in this study. The main contributions are summarized as follows: 
\begin{itemize}
    \item 
    Our model quantifies the impact of predictive information on resilience planning, underscoring its criticality. Specifically, two distinct aspects are incorporated into the RIDDRP model: the time-varying accuracy and the anticipatory preparation time prior to ice storms. This integration allows for more optimized utilization of available resources to enhance system resilience.
    \item Considering the changes in operation objectives under normal conditions in anticipating ice storms and emergency conditions during ice storms, our model emphasizes risk-informed coordination between those conditions to achieve judicious investment decisions. A dual-objective approach is employed to achieve trade-offs between economy and resiliency. Thus, avoiding the over-design of the system and saving unnecessary costs.
     
    \item The inherent uncertainty of predictive information, i.e., the time-varying accuracy and the associated timing of anticipatory preparations,
    the DDU introduced by line hardening decisions, and various ice storm-related exogenous uncertainties are incorporated and formulated. By addressing these multifaceted uncertainties, our model can characterize their coupled effect on the system operation under ice storms to avoid potential insecure issues.
    
    \item A two-stage risk-informed decision-dependent resilience planning (RIDDRP) model with mixed-integer recourse is constructed. The first stage makes line hardening decisions, as well as the optimal sitting and sizing of energy storage. The second stage evaluates the risk-informed operation costs.
    To relieve computational burdens, the progressive hedging algorithm (PHA) 
    is adopted to solve the large-scale mixed-integer linear programming problem. To validate the efficacy of the proposed model, comparative analyses are conducted under various anticipatory preparation times. 
    It is shown that the proposed resilience enhancement strategy can bolster system resilience enhancement by adopting more judicious investment
    and optimizing the utilization of dispatchable resources.
\end{itemize}

The remainder of this paper is organized as follows: Section II introduces the system modelling under ice storms. Section III quantifies the impact of predictive information. Section Iv describes the mathematical formulation of the proposed two-stage RIDDRP model and the solution algorithm. Section IV discusses the solving algorithms as well as the lower bound guarantee. Section V presents case studies and Section VI concludes this study. 

\vspace{-0.5em}
\section{The Fragility Model of Transmission System Under Ice Storms}
\subsection{Ice storms-induced System Outage}
In addressing power transmission system vulnerabilities under various weather intensities, we adopt the concept of the fragility curve to model the relationship between the failure probability of system components and the weather intensity. Under ice storms, the weather intensity $z$ is denoted by the ice thickness of ice $r$. Given the meteorological data, a freezing rain ice accretion model is utilized to calculate ice loads \cite{lu2018resilience}:
\begin{equation}
\label{Iceaccretion}
r = \frac{h}{\rho_{i}\pi} \sqrt{(P\!r\rho_{w})^{2}+(3.6vL)^{2}}
\end{equation}
where $r$ represents the ice thickness, $Pr$ and $L$ denote the precipitation rate and the liquid water content, respectively. The following empirical formulation is utilized to model their relationship:
\begin{equation}
\label{Iceaccretion2}
L = 0.067Pr^{0.846}
\end{equation}
where $h$ denotes the hours of freezing. $\rho_i$ and $\rho_w$ are the density of ice and water with $0.9g/cm^3$ and $1g/cm^3$, respectively. $v$ is the wind speed.
Historical statistic analysis can be conducted to estimate proper probability distribution functions (PDFs) for each weather-related parameter. In this study, PDFs stated in \cite{abdelmalak2021resilience} are adopted to model ice storms. 

As an ice storm propagates through the transmission system, various transmission corridors will be affected. 
To simplify the problem, we assume that all lines share the same fragility curve. The failure probabilities are independent of each other for a fixed weather intensity. Then, the failure probability of a component can be expressed as:
\begin{equation} 
\label{piecewiselinear}
P_f(r)=
\left\{
\begin{aligned}
&0 \qquad \qquad \qquad \qquad \qquad \  \ \ r < R
\\\
&{\rm exp}\left[\frac{0.6931(r-R)}{4R}\right]-1 \ \ R \le r < 5R
\\\
& 1 \qquad \qquad \qquad \qquad \qquad \quad r \ge 5R
\end{aligned}
\right.
\end{equation}
where $R$ is determined by the line hardening decision $\boldsymbol{x}_{ij}$.

Considering the spatial and temporal changes of ice storms, 
a transmission corridor is divided into $L$ line segments and a failure in a single segment will result in the failure of the entire corridor due to their connection in series. The total duration of the ice storm $T$, is divided into $N$ time steps with a shorter duration period $\Delta t$, Then, the failure probability of the $k$-th segment in the $i$-th corridor at time $t_j$ can be evaluated once given the meteorological data.
Due to the series connection, the cumulative failure probability of the transmission corridor connecting bus $i$ and $j$ during the ice storm period $T$ can be evaluated using all its segments:
\begin{equation} 
\label{cumulativeFailure}
P_{ij}^f(r)=1-\prod \limits_{l=0}^L (1-P_{fl}(r))
\end{equation}
where $P_{fl}(r)$ is the failure probability of the $l$-th segment and $P_{i}(r)$ is the cumulative failure probability of the $i$-th transmission corridor.

\subsection{Decision-dependent Line Failure Probability} 
In this study, the endogenous uncertainty lies in the realization of contingencies influenced by the investment decisions made in the planning stage. Specifically, the ability of a transmission line to resist ice accretion can be manipulated through purposeful hardening designs. Consequently, the uncertain line damage status $\mu_{ij,t}$ under a certain ice storm scenario is interrelated with the line hardening decisions and becomes DDU. 

In this model, the empirical fragility function is adopted to estimate the failure probability of a given transmission line $P_f$, which is parameterized by both the endogenous hardening decisions $\boldsymbol x$ and the exogenous intensity of ice storms. Under the same ice thickness, the failure probability of a transmission line will decrease if it is hardened. Since the potential number of $\boldsymbol x$ will grow exponentially with the size of the system, it would be unrealistic to generate scenario sets that contain all possible realizations of $\boldsymbol x$. To tackle this problem, we decompose the outcome of a random event $\mu_{ij,t,s}$, i.e., whether the line is damaged by ice storms, into two conditional independent binary variables $\phi_{ij,t,s}^w$ and $\phi_{ij,t,s}^o$, which represent the line damage status with or without line hardening respectively. The damage status variables will be set as 0 until line damage occurs.
\begin{equation} 
\label{lineDDU}
\mu_{ij,t,s} = (1-x_{ij,s})\phi_{ij,t,s}^o+x_{ij,s}\phi_{ij,t,s}^w
\end{equation}

Thus, the line status under a certain scenario $\mu_{ij,t,s}$ can be represented as an explicit function of the first-stage line hardening decisions $\boldsymbol x$. Then, $\phi_{ij}^w$ and $\phi_{ij}^o$ can be determined in advance during the scenario generation process to handle the uncertainty of line damage status. Therefore, we can proactively mitigate line damage risk through strategic hardening decisions.

\begin{figure}[!t]
\centering
\includegraphics[scale=0.7]{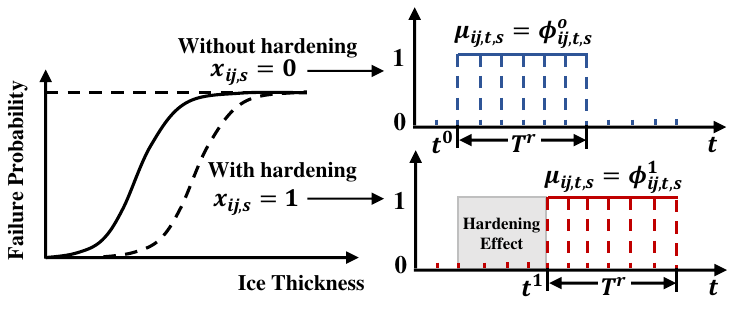}
\vspace{-0.4em}
\caption{Decision-dependent damage status of line $ij$ with/without hardening 
decision.}
\vspace{-0.6em}
\label{DDU}
\end{figure}

To model the inter-temporal correlation during the repair process, the line status is treated as a Bernoulli process and we take the first time stamp $t^{0/1}$ that $\phi_{ij,t,s}^{o/w} = 1$ as the initiation of line damage, which can be sampled from the fragility curve \ref{piecewiselinear}. During the repairing process that lasts $T^r$ hours, the line status will be fixed as 1 to represent line outages. If the line is repaired before the end of the ice storm, we assume that this line will not be damaged again in the remaining time of the ice storm. The whole process of determining the line status $\mu_{ij,t,s}$ can be illustrated in Fig. \ref{DDU}. 

By incorporating DDU during planning, we are able to incorporate not only the exogenous intensity of ice storms but also the endogenous line strength to withstand certain levels of ice accretion, thereby further releasing the system's potential for resilience enhancement in the face of extreme events.

\vspace{-1.3em}
\subsection{Ice Storm-related Exogenous Uncertainties}

\subsubsection{Uncertainty of wind turbine blade freezing}  
During ice storms, ice accumulation can lead to the failure of wind turbines. In our model, the failure probability of a wind turbine follows log-logistic distribution for a given ice thickness $r$:
\begin{equation}
\label{WTfailure}
P_{WT}^f(r) = \frac{(r/\alpha)^\beta}{1+(r/\alpha)^\beta}
\end{equation}
where $\alpha$ and $\beta$ are both parameters of the log-logistic distribution that can be estimated using historical data \cite{rose2013quantifying}. 




\subsubsection{Uncertainty of Load}
Given the historical load profile $L^H$, we assume a homogeneous load pattern prior to and during the ice storm. The load uncertainty level $\kappa_{i,s}$ that follows a normal distribution is utilized to generate load profiles for a stochastic scenario \cite{ma2016resilience}:
\begin{equation} 
\label{LoadU}
Load_{i,t,s}^{D} = \kappa_{i,s}L^H
\end{equation}

\subsubsection{Uncertainty of repair time}
The repair time of each damaged component $T^{r}(s)$, is a random variable that depends on dispatches of repair crews and resources. In this study, we assume that the repair time is independent and follows the same Weibull distribution \cite{arab2015stochastic}.  The damaged status of the line or wind turbine will last until the end of the repair time.
For each damaged line $ij$ or wind turbine $w$, the repair time is:
\begin{equation} 
\label{Weilbull}
T^{r}_{ij/w}(t)=
\left\{
\begin{aligned}
& \!\! \frac{\beta}{\alpha}
\!\left(\frac{t}{\alpha}\right)
^{\!\beta-1}
\!
\!\!\!\! {\large e}^{-t^\beta / \alpha^\beta} \quad \ t \ge 0
\\\
& \!\! 0  \qquad \qquad \qquad \qquad \rm otherwise
\end{aligned}
\right.
\end{equation}
where $\alpha=4$ and $\beta=10$ are parameters of the distribution. 

\section{Quantification of the impact of Predictive Information}
In this section, we delve into the critical role of predictive information. Specifically, our focus is on quantifying the impact of predictive information through two key concepts: the time-varying accuracy and the anticipatory preparation time. The former aims to model the evolving accuracy, and the latter aims to model its operational impact over time. By mathematically modeling the impact of predictive information, we aim to integrate those characteristics into the decision-making process for resilience enhancement.  

\subsection{Time-varying accuracy quantification}

\begin{figure}[!t]
\centering
\includegraphics[scale=0.5]{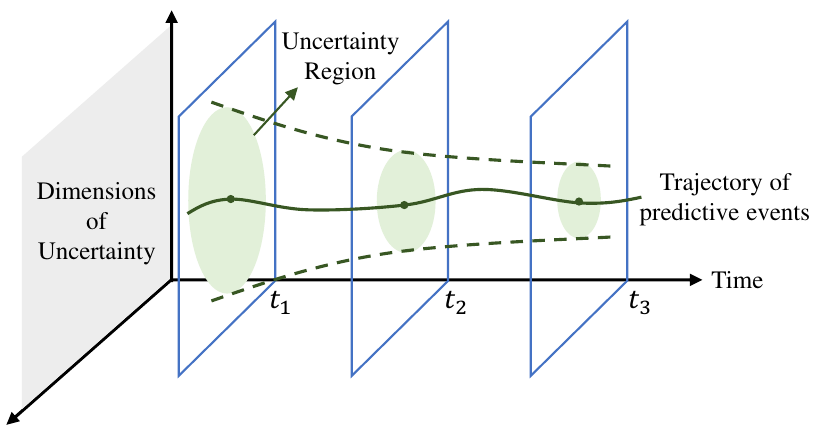}
\vspace{-0.5em}
\caption{Illustrative example of the uncertainty level evolution in predictive events.}
\label{UncertainTraj}
\vspace{-1.2em}
\end{figure}

Given the ability to forecast ice storms, operators can obtain critical predictive information prior to these events, enabling proactive preparation. However, the inherent uncertainty in the predictive information, such as the forecasting accuracy, has seldom been considered and properly modeled. 
By incorporating time-sensitive information, the uncertainty and availability of resources can be considered in the model, 
which can significantly influence the dispatch of hybrid resources for better preparation and response against ice storms. 
This, in turn, can lead to the development of more effective operation strategies adapted to the evolving circumstances.
As the event approaches, the uncertainty facilitating proactive preparation tends to decrease, enabling more precise preparatory measures. This phenomenon can be mathematically encapsulated by a monotonically decreasing characteristic. 
\begin{equation}
\label{UncertaintyDecrease}
\mathcal{U}(X_{t+1}) \le \mathcal{U}(X_t), \ \forall t
\end{equation}
where $X_t$ denotes the predictive information at time $t$ and $\mathcal{U}$ represents the uncertainty level of the current information. 
The associated confidence intervals can be expressed as:
\begin{equation}
\label{CI}
CI(X_t)=[\mu_t-z\sigma_t, \mu_t+z\sigma_t]
\end{equation}
where  $\mu_t$ and $\sigma_t$ are the mean and standard deviation of the predictive information at time $t$, and $z$ is the z-score corresponding to the desired confidence level. The decreasing tendency of the uncertainty associated with the predictive information can be characterized by the gradually narrowed confidence intervals $CI(X_t)$ as time approaches the occurrence of the anticipated events. An illustrative example of the evolutionary trajectory of uncertainty levels in predictive events is shown in Fig.\ref{UncertainTraj}.

Although numerous posterior indices exist to quantify the accuracy of predictive information, preemptively estimating this accuracy remains a challenge.  In this study, we navigate this complexity by mapping the uncertainty to a time-varying preventive load-shedding penalty, denoted as $c_{LS}^N$, before the onset of ice storms. This can be achieved through a generally monotonically decreasing function, which serves to quantify the progressively diminishing uncertainty of predictive information as the event approaches. This approach is grounded in the rationale that initiating proactive measures at an earlier stage can be riskier since there is higher uncertainty for future conditions. The decisions made under current conditions may not remain valid or advantageous in the future, thereby elevating the associated risks and potential costs of load shedding.
It is noteworthy that other types of functions can also be selected based on the specific problem at hand. 

\subsection{Anticipatory preparation time}

As mentioned above, predictive information enables preventive strategies against ice storms. However, differences in operational strategies under normal and emergency conditions should be acknowledged. Traditional planning objectives aim at ensuring resource adequacy, which can lead to resource depletion during emergencies. In contrast, the objective of resilience planning requires the consideration of both normal and emergency states. Under normal conditions, the objectives align with traditional planning, focusing on resource adequacy. However, during emergencies, the objective shifts to maximizing the critical load that can be retained by coordinating resources across different stages. The change of operational objectives has seldom been considered in past research. 


To present a more holistic and robust planning strategy, in this study, both the normal operation conditions prior to ice storms and the emergency conditions during extreme events are incorporated as the planning objective to strike a balance between economic efficiency and system resiliency. 
A critical aspect of this strategy is the concept of \textit{anticipatory preparation time}, which refers to the period before an ice storm when preparations are initiated based on predictive information. 
Given the available predictive information and its associated uncertainty level, operators can select an appropriate time to start preparations for anticipated ice storms. This decision is crucial as it allows for the strategic pre-dispatch of resources, including charging energy storage, to mitigate the risk of large-scale load shedding during ice storms. The determination of this commencing time is influenced not only by the need for operational strategy adjustments but also by the capabilities of current forecasting technology.

Due to the time-varying uncertainty associated with predictive information, which we have mapped to a preventive load-shedding penalty that decreases with time, the anticipatory preparation time is directly linked to the economic risks. The earlier the preparation anticipatory preparation time, the higher the economic risks.
Therefore, operators must judiciously choose the start time for these preparations, balancing the benefits of early action against the escalating economic risks posed by prolonged periods of uncertainty.
In this study, the anticipatory preparation time is parameterized by $\xi$, representing the time interval between the initiation of risk-informed operation and the occurrence of the ice storms. Various values of \(\xi\) will be selected in the case study to compare and explore the impact of this parameter on system planning and operations.
Considering the transition from normal conditions to extreme events, the set of normal operating conditions and the emergency operating conditions is determined by the advanced time determined by the predicted information. Since the duration of extreme events is an exogenous factor affected by nature, $T^{E}$ can be viewed as a fixed set. The set of risk-informed normal operation time $T^{N}$ duration is parameterized by the advanced time $\xi$ influenced by the technology level and the operator's decisions \cite{degelia2016overview}. Thus, there is:
\begin{equation} 
T^{N}=\{t_{t}|i=t_{E}\!-\!\xi,\dots,t^{E}_{\rm start}\!-\!2,t^{E}_{\rm start}\!-\!1\}
\end{equation}
\begin{equation}
\label{Psi2}
T^{E}=\{t_{t}|i=t^{E}_{\rm start},\dots,t^{E}_{\rm end}\}
\end{equation}

Subsequently, we adopted a dual objective to model the changes in the operational strategies:
\begin{equation}
\label{Dualobj}
\psi_s(T) = \psi^{\rm N}_s(T^{N}) +\psi^{\rm E}_s(T^{E})
\end{equation}
where $\psi_s$ is the total cost given a specific scenario $s$, $\psi^{\rm N}_s$ and $\psi^{\rm E}_s$ are the associated normal and emergent operational cost, respectively. Given the circumscribed anticipatory preparation time, this approach empowers operators to dispatch available resources more efficiently, thereby furnishing insights during the planning stage.


\section{Mathematical Formulation of The two-stage risk-informed resilience planning model}
\begin{figure}[!t]
\centering
\includegraphics[scale=0.5]{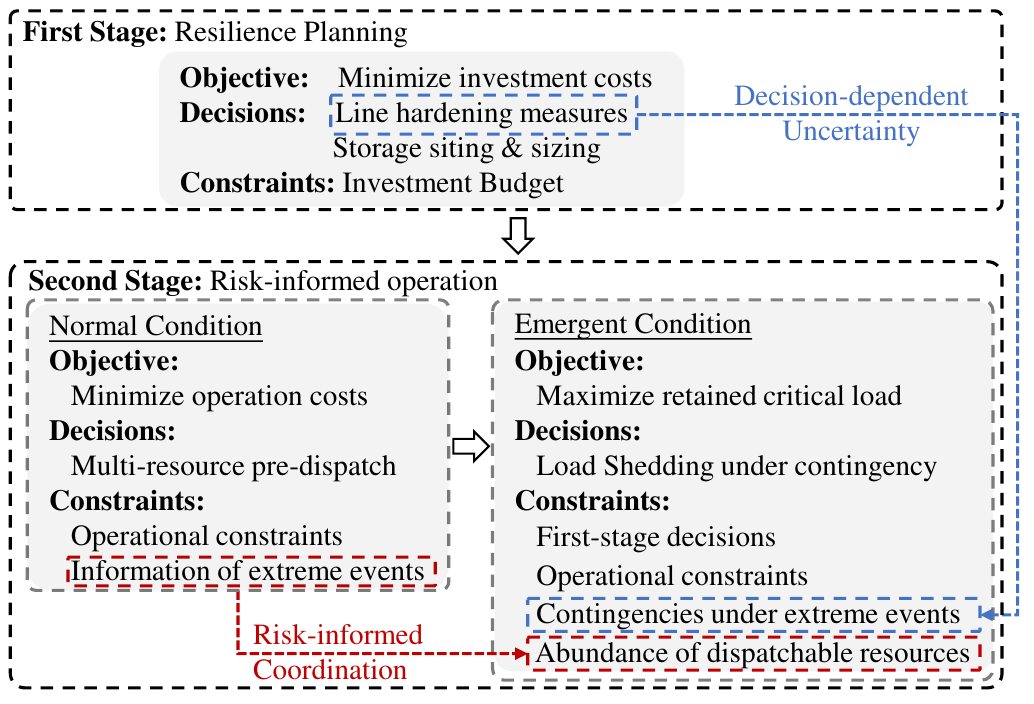}
\vspace{-0.4em}
\caption{The proposed two-stage risk-informed resilience planning model.}
\vspace{-0.6em}
\label{RPFrame}
\end{figure}

In this section, we propose a two-stage stochastic mixed-integer programming model for risk-informed resilience planning considering ice storms, as shown in Fig.\ref{RPFrame}.
The first stage is to make system planning decisions, i.e., line hardening as well as sitting and sizing of energy storage. The second stage is for operational cost evaluation under ice storm scenarios, which contains normal conditions prior to and emergent conditions during ice storms. By jointly optimizing the entire operational cost before and during the events, we aim to achieve the trade-off between costs under different scenarios to enhance system resilience while maintaining economic efficiency. 

\vspace{-1em}
\subsection{Objective Function}
The object function of the proposed two-stage stochastic resilience planning model is as follows:
\begin{equation}
\label{obj}
\text{min} \quad \sum_{l \in L} c_{l}x_{l} + \sum_{i \in N} c_{b}Z_{i} + \mathbb{E}_{s}[Q(\boldsymbol{x},\boldsymbol{z},s)]
\end{equation}

The first and second terms of \eqref{obj} are the investment cost of line hardening and energy storage, respectively. The third term is the expected value under normal and extreme operation, respectively. The mathematical representation of this term is as below:
\begin{equation}
\mathbb{E}_{s}[Q(\boldsymbol{x},\boldsymbol{z},s)] = \sum_{s \in S} p_{s}(\psi^{\rm N}(\boldsymbol{x},\boldsymbol{z},s) +\psi^{\rm E}(\boldsymbol{x},\boldsymbol{z},s))
\label{expectation}
\vspace{-1em}
\end{equation}
where
\begin{equation}
\label{Psi1}
\begin{aligned} 
\psi^{\rm N}(\boldsymbol{x},\boldsymbol{z},s) = \text{min}\sum_{t \in T^{N}}(\sum_{i \in N} c_{g}P^{g,N}_{i,t,s} + \sum_{i \in N} c_{dis}P^{dis,N}_{i,t,s} 
\\
+ \sum_{i \in N}c_{w} P_{i,t,s}^{wc,N} + \sum_{i \in N}c_{LS}^N P^{LS,N}_{i,t,s})  
\end{aligned}
\end{equation}
\begin{equation}
\label{PsiE}
\psi^{\rm E}(\boldsymbol{x},\boldsymbol{z},s) = \text{min} \!\sum_{t \in T^{E}} \sum_{i \in N} c_{LS}^E P^{LS,E}_{i,t,s}
\end{equation}

The three terms in \eqref{Psi1} aim to minimize the operation costs of generators, storage, wind curtailment, and load shedding under normal conditions, respectively. Equation \eqref{PsiE} is to minimize the cost of load shedding under emergency conditions, i.e., ice storms. 

It should be noted that load shedding is introduced in the normal operation stage as a proactive measure to mitigate the impact of anticipated extreme events, which are leveraged by predictable information. The rationale is that pre-emptive charging of storage may introduce an imbalance between generation and demand. Therefore, preventive load shedding and storage charging are necessary to deal with potential power disruption. Since cost-effectiveness is still the major objective during normal operation, the coefficient $c_{LS}^N$ can be adjusted to achieve the trade-off between economic efficiency and system resilience.

\vspace{-0.5em}
\subsection{Constraints}
In this two-stage stochastic mixed-integer linear programming model, the constraints can be divided into three parts, including investment budgets, operational constraints under normal conditions and extreme events, respectively. 

In the first stage, the investment decisions are made according to related budgets, which are given as follows:
\begin{equation}
\label{LineBudget}
\sum_{l \in L} c_{l}x_{l} \le C^{L}
\end{equation}
\begin{equation}
\label{BatteryBudget}
\sum_{i \in N} c_{b}Z_{i} \le C^{Z}
\end{equation}
\begin{equation}
\label{LineHardenDecision}
x_{l}\in \{0,1 \}, \forall l \in L
\end{equation}

In the second stage, there are two types of constraints. One is for normal operation prior to extreme events, the other is for emergent operation during ice storms. Since there is an anticipation of extreme events and the anticipatory preparation time given by the predictive information is represented by $T^{N}$, operators are capable of planning preparation methods such as multi-resource re-dispatch. The operational constraints under normal conditions are as follows:
\begin{equation}
\label{NormalPowerFlow}
\begin{aligned}
P^{g,N}_{i,t,s} + \sum_{ji \in L}\!\!\!P_{ji,t,s}^{N} - \sum_{ij \in L}\!\!\!P_{ij,t,s}^{N} - (P_{i,t,s}^{D,N} - P^{LS,N}_{i,t})
\\ + P_{i,t,s}^{w,N} - P_{i,t,s}^{wc,N} + P_{i,t,s}^{ch,N} - P_{i,t,s}^{dis,N} = 0
\end{aligned}
\end{equation}
\begin{equation}
\label{PFN}
P_{ij,t,s}^{N} + P_{ji,t,s}^{N} = B_{ij}(\theta_{i,t,s}-\theta_{j,t,s})
\end{equation}
\begin{equation}
\label{LineNCons}
-P_{ij}^{\rm Max} \le P_{ij,t,s}^{N} + P_{ji,t,s}^{N} \le P_{ij}^{\rm Max}
\end{equation}
\begin{equation}
\label{GenConstN}
P^{\rm Min}_{g} \le P^{g,N}_{i,t,s} \le P^{\rm Max}_{g}
\end{equation}
\begin{equation}
\label{LoadShedN}
0 \le P^{LS,N}_{i,t} \le \alpha_i P_{i,t,s}^{D,N}
\end{equation} 
\begin{equation}
\label{WCN}
0 \le P_{i,t,s}^{wc,N} \le P_{i,t,s}^{w,N}
\end{equation}
\begin{equation}
\label{Dis&ChN}
x^{dis,N}_{i,s} +  x^{ch,N}_{i,s} \le 1 
\end{equation}
\begin{equation}
\label{Dis&ChNB}
x^{dis,N}_{i,s},x^{ch,N}_{i,s} \in \{0,1\}
\end{equation}
\begin{equation}
\label{DisN}
0 \le P^{dis,N}_{i,t,s} \le x^{dis,N}_{i,s} P^{\rm Max}_{b}
\end{equation}
\begin{equation}
\label{ChN}
0 \le P^{ch,N}_{i,t,s} \le x^{ch,N}_{i,s} P^{\rm Max}_{b}
\end{equation} 
\begin{equation}
\label{SOCN}
S_{i,t-1,s}^{N} - S_{i,t,s}^{N} = (P^{dis,N}_{i,t,s}/\eta^{+} - P^{ch,N}_{i,t,s}/\eta^{-}) \cdot \Delta{t}
\end{equation}
\begin{equation}
\label{SN}
0 \le S_{i,t,s}^{N} \le Z_{i} 
\end{equation}
\begin{equation}
\label{PtoEratioN}
P^{\rm Max}_{b}\rho_{N} \le Z_{i} 
\end{equation}

where (\ref{NormalPowerFlow}) describes the power balance between supply and demand, (\ref{PFN}) and (\ref{LineNCons}) represent the power flow equations and line flow limits, (\ref{GenConstN}) represents the generator output limits and (\ref{LoadShedN}) sets the 
 tolerable load shedding ratio of each bus and the critical loads usually have lower tolerance of load shedding, (\ref{WCN})
sets the wind curtailment limits, (\ref{Dis&ChN}) and (\ref{Dis&ChNB}) describe that the energy storage can not charge and discharge simultaneously, (\ref{DisN}) and (\ref{ChN}) set the discharging/charging limits of energy storage, (\ref{SOCN}) models the dynamic equation of energy storage. 
(\ref{SN}) sets the state-of-charge limit, and (\ref{PtoEratioN}) relates the energy and power ratings of storage. 
Similarly, the operational constraints during extreme events are as follows:
\begin{equation}
\label{EmergPowerFlow}
\begin{aligned}
P^{g,E}_{i,t} + \sum_{ji \in L}\!\!\!P_{ji,t,s}^{E} \!-\! \sum_{ij \in L}\!\!\!P_{ij,t,s}^{E} - (P_{i,t,s}^{D,E} \!-\! P^{LS,E}_{i,t})
\\ + \gamma_{i,t,s}(P_{i,t,s}^{w,E} - P_{i,t,s}^{wc,E}) + P_{i,t,s}^{ch,E} \!-\! P_{i,t,s}^{dis,E} = 0
\end{aligned}
\end{equation}
\begin{equation}
\label{ExtremeFlowLimit}
- (1 - \mu_{ij,t,s}) P_{ij}^{\rm Max}\!\le\!P_{ij,t,s}^{E} + P_{ji,t,s}^{E}\!\le\!(1-\mu_{ij,t,s}) P_{ij}^{\rm Max}
\end{equation} 
\begin{equation}
\label{ExtremeFlowLimit2}
\begin{aligned}
-\mu_{ij,t,s}M\!\!\le\!\!P_{ij,t,s}^{E}\!\!+\!P_{ji,t,s}^{E}\!\!-\!B_{ij}\!(\theta_{i,t,s}\!-\!\theta_{j,t,s}\!)\!\!\le\!\mu_{ij,t,s}M 
\end{aligned}
\end{equation}
\begin{equation}
\label{LineECons}
-P_{ij}^{\rm Max} \le P_{ij,t,s}^{N} + P_{ji,t,s}^{N} \le P_{ij}^{\rm Max}
\end{equation}
\begin{equation}
\label{GenConstE}
P^{\rm Min}_{g} \le P^{g,E}_{i,t,s} \le P^{\rm Max}_{g}
\end{equation}
\begin{equation}
\label{GenConst}
0 \le P^{LS,E}_{i,t,s} \le \beta_i P_{i,t,s}^{D,E}
\end{equation} 
\begin{equation}
\label{WindConst}
0 \le P_{i,t,s}^{wc,E} \le \gamma_{w,t,s}P_{i,t,s}^{w,E}
\end{equation}
\begin{equation}
x^{dis,E}_{i,t,s} +  x^{ch,E}_{i,t,s} \le 1 
\end{equation}
\begin{equation}
x^{dis,E}_{i,t,s},x^{ch,E}_{i,t,s} \in \{0,1\}
\end{equation}
\begin{equation}
\label{ChConst}
0 \le P^{ch,E}_{i,t} \le x^{ch,E}_{i,t,s} P^{\rm Max}_{b}
\end{equation}
\begin{equation}
\label{DisConst}
0 \le P^{dis,E}_{i,t} \le x^{dis,E}_{i,t,s} P^{\rm Max}_{b}
\end{equation}
\begin{equation}
S_{i,t-1,s}^{E} - S_{i,t,s}^{E} = (P^{dis,E}_{i,t,s}/\eta^{+} - P^{ch,E}_{i,t,s}/\eta^{-}) \cdot \Delta{t}
\end{equation}
\begin{equation}
0 \le S_{i,t,s}^{E} \le Z_{i}
\end{equation}
\begin{equation}
\label{PtoEratioE}
P^{\rm Max}_{b}\rho_{E} \le Z_{i}
\end{equation}

In the above constraints, the binary variables $\gamma_{i,t,s}$ and $\mu_{ij,t,s}$ denote the 
damage status of wind farms and transmission lines, respectively. If the wind farm $i$ is out-of-service at time interval $t$ in scenario $s$, then $\gamma_{i,t,s}=1$, otherwise $\gamma_{i,t,s}=0$. Similarly, if line $ij$ is damaged at time interval $t$ in scenario $s$, then $\mu_{ij,t,s}=1$, otherwise $\mu_{ij,t,s}=0$.

The occurrence of predictable information can influence the normal operation before extreme events to make preparations. The system states at the beginning of the extreme events, or the end of the normal operation, will be different given different preparation times, which can be denoted as:
\begin{equation}
\label{Dis&ChConst}
S_{i,t_{0},s}^{E} = S_{i,t_{n},s}^{N}
\end{equation}
 

\textbf{\textit{Remarks:}} While some research has considered operation under normal and emergency conditions, the distinctive feature of our model lies in the integration of predictive information and the coupling between the two stages. Instead of being treated as independent scenarios, preventive normal operation (\ref{NormalPowerFlow})-(\ref{PtoEratioN}) and emergent operation (\ref{EmergPowerFlow})-(\ref{PtoEratioE}) collectively form a comprehensive scenario through the explicit coupling of (\ref{Dis&ChConst}) and the implicit coupling across consecutive time intervals. Therefore, our model emphasizes the coordination both prior to and during contingencies, influenced by the anticipatory preparation time, which is determined by predictive information and its associated time-varying uncertainty.

\vspace{-0.5em}
\subsection{Scenario Generation}
In general, the sample space of the stochastic programming problem described in (\ref{obj})-(\ref{Dis&ChConst}) is of infinite dimensions. To cope with this difficulty, we consider finite realizations of scenarios to construct an
approximation of the resilience planning problem. The scenario set $S$ is restricted to denote the set of scenarios $s$ with corresponding probabilities $p_s$. The scenario set is generated by applying the general procedure stated in Algorithm 1.


\renewcommand{\algorithmicfor}{\textbf{For}}
\begin{algorithm}[h] 
\caption{Scenario Generation} 
\begin{algorithmic}[1]
\setstretch{1.}
\State \textbf{Input:} Parameters $T^{N}$, $T^{E}$ and weather-related data.
\For{scenario $s=1,...,|S|$}
    \State Sample ice storm parameters using the related PDFs. 
    \State Sample $\kappa_{i,s}$ from $N(1, 0.1)$, $\forall i \in \Omega_L$ 
    \State Generate load profile via (\ref{LoadU})
    \For{$ij \in \Omega_{B}$}
        \State Sample $T^{r}_{ij}(s)$ via (\ref{Weilbull})
        \State Calculate $r(s)$ via (\ref{Iceaccretion})
    \State Calculate $P_{ij}^f(r)$ and $P_{WT}^f(r)$ via (\ref{cumulativeFailure}) and (\ref{WTfailure})
    \State Sample component status $\mu_{ij}(s)$ and $\gamma_{w}(s)$
    \State $ 
    \mu_{ij,t} / \gamma_{w,t}(s)\!=\!
    \left\{
    \begin{aligned}
    &1\ \  \text{min}\{t\!+\!T^{r}_{ij/w}(s)\!-\!1, T\} \!\le\! t \!\le\! T
    \\
    &0\ \  \text{otherwise}
    \nonumber
    \end{aligned}
    \right. $
    \renewcommand{\algorithmicfor}{\textbf{for}}
    \EndFor
\EndFor
\end{algorithmic}
\label{alg1}
\end{algorithm}


\subsection{Solution Algorithm}
A large number of scenarios result in extremely large-scale MILP resilience planning models.
Commonly used stage-wise decomposition methods, such as Branch and Bound, can be computationally intensive, especially for large-scale problems with binary variables \cite{zhang1996branch}. Benders’ decomposition, which is another general method for solving large-scale mixed-integer stochastic programming problems, requires the convexity of $\psi(\boldsymbol{v},s)$ and becomes invalid when there are integer variables in the second stage.
To mitigate the computational difficulty, the progressive hedging algorithm (PHA) is adopted to solve the large-scale RDDIP model. PHA first decomposes the extensive form of the original problem according to each scenario. Then, the penalized versions of the subproblems will be solved parallelly and iteratively until convergence. 

To implement the PHA, the RDDIP model is rewritten in the following compact form:
\begin{align} 
\label{Compactobj1}
\mathop{{\rm min}}\limits_{\boldsymbol{v}}& \quad \boldsymbol{c}^{\top} \boldsymbol{v} + \sum_{s \in S} p(s)\psi(\boldsymbol{v},s) \\
\label{Compactcons1}
{\rm s.t.}& \quad \boldsymbol{Av} \le \boldsymbol{B} \\
\label{decision1}
& \quad \boldsymbol{v} \in \mathbb{Z}^{m_1}_+\times\mathbb{R}^{n_1-m_1}
\end{align}
where $\boldsymbol{v}$ represents the binary line hardening decision variables and the storage siting and sizing decision variables in the first stage, as shown in (\ref{obj}). $\boldsymbol{c}$ is the corresponding cost coefficient vector. (\ref{Compactcons1}) is the matrix form of the investment budget constraints as shown in (\ref{LineBudget}) - (\ref{LineHardenDecision}). $\psi(\boldsymbol{v},s)$ represents the second-stage operation problem under scenario $s$, which can be expressed as:
\begin{align} 
\label{Compactobj2}
\psi(\boldsymbol{v},s) = \ {\rm min}& \ \ \boldsymbol{d}^{\top}(s) \boldsymbol{q} \\
\label{Compactcons2}
{\rm s.t.}& \ \ \boldsymbol{Dq} \le \boldsymbol{F}(s)-\boldsymbol{H}(s)\boldsymbol{v} \\
\label{decision2}
& \quad \boldsymbol{q} \in \mathbb{Z}^{m_2}_+\times\mathbb{R}^{n_2-m_2}
\end{align}
where $\boldsymbol{q}$ represents the second-stage decision variables. (\ref{Compactcons2}) is the matrix form of the operational constraints corresponding to (\ref{NormalPowerFlow}) - (\ref{Dis&ChConst}). 

To render the problem solvable, we construct an approximation of the problem by considering finite scenarios $|S|$. Then, we can obtain the scenario formulation of the RDDIP model: 
\begin{align} 
\label{SF1}
\mathop{{\rm min}}& \quad \sum_{s \in S}[\boldsymbol{c}^{\top} \boldsymbol{v}(s) +  p(s)\boldsymbol{d}(s)^{\!\top} \boldsymbol{q}(s)] \\
\label{SF2}
{\rm s.t.}& \quad \boldsymbol{Av} \le \boldsymbol{B} \\ 
\label{SF3}
& \quad \boldsymbol{Dq} \le \boldsymbol{F}(s)-\boldsymbol{H}(s)\boldsymbol{v} \\
\label{SF4}
& \quad \boldsymbol{v} - \boldsymbol{\hat{v}} =0 \\
\label{SF5}
& \quad \boldsymbol{v},\boldsymbol{\hat{v}} \in \mathbb{Z}^{m_1}_+\times\mathbb{R}^{n_1-m_1} \\
\label{SF6}
& \quad \boldsymbol{q} \in \mathbb{Z}^{m_2}_+\times\mathbb{R}^{n_2-m_2}
\end{align}
where $\boldsymbol{\hat{v}}$ is the copy of the first-stage variables for
each scenario. (\ref{SF4}) is the so-called non-anticipativity constraints, which stipulate that in all feasible solutions, the first-stage decisions are independent on scenarios. Without (\ref{SF4}), the extensive form of the scenario formulation can be decomposed by scenario.  

The PHA is initialized by decomposing the large-scale, two-stage RDDIP model into scenario-specific subproblems, each of which is then solved independently. In each iteration, the solution of the individual scenario solutions will be projected onto the subspace of non-anticipative constraints for aggregation. Non-anticipativity is enforced by using penalties for multiplier updates. Then, each subproblem whose first-stage objectives are perturbed by the multipliers will be solved. 
The procedure of implementing the PHA to solve the two-stage stochastic MILP resilience planning problem is shown in Algorithm 2. 

\renewcommand{\algorithmicwhile}{\textbf{While}}
\begin{algorithm}[t]
\setstretch{1.}
\caption{The Progressive Hedging Algorithm for Two-Stage Stochastic MILP Resilience Planning Problems} 
\begin{algorithmic}[1]
\vspace{0.05cm}
\State \textbf{Initialization:} 
\Statex \hspace{0.2cm} Let $k \leftarrow 0$ and $w^k(s) \leftarrow 0$. For each $s \in S$ calculate:
\Statex \hspace{0.2cm} $(\boldsymbol{v}^{k+1}(s),\boldsymbol{q}^{k+1}(s)) \in \! \mathop{\rm argmin}\limits_{(\boldsymbol{v},\boldsymbol{q}) \in \Psi(s)}
\boldsymbol{c}^{\!\top}\boldsymbol{v} +\boldsymbol{d}(s)^{\!\top} \boldsymbol{q}$
\vspace{0.1cm}
\Statex \hspace{0.2cm} $\Psi(s) := \{\boldsymbol{v} \in \mathbb{Z}^{m_1}_+\times\mathbb{R}^{n_1-m_1},\boldsymbol{q} \in \mathbb{Z}^{m_2}_+\times\mathbb{R}^{n_2-m_2}\!:$
\Statex \hspace{1.6cm} $\boldsymbol{Av} \le \boldsymbol{B}, \boldsymbol{Dq} \le \boldsymbol{F}(s)-\boldsymbol{H}(s)\boldsymbol{v}\}$
\While{all scenario solutions $\boldsymbol{v}^k(s)$ are unequal}
    \State \textbf{Iteration Update:} $k \leftarrow k+1$
    \State \textbf{Aggregation:} $\boldsymbol{\hat{v}}^k \leftarrow \sum \limits_{s\in S}p(s)\boldsymbol{v}^k$
    \State \textbf{Multiplier Update:} $w^k(s) \! \leftarrow \! w^{k-1}(s)+\rho(\boldsymbol{v}^k(s)-\boldsymbol{\hat{v}}^k)$ 
    \State \textbf{Decomposition:} 
    \Statex \hspace{0.8cm} For each $s \in S$, calculate: $(\boldsymbol{v}^{k+1}(s),\boldsymbol{q}^{k+1}(s)) \in$
    \Statex \hspace{0.7cm} $ \mathop{\rm argmin}\limits_{(\boldsymbol{v},\boldsymbol{q}) \in \Psi(s)} \{\boldsymbol{c}^{\top} \boldsymbol{v} +\boldsymbol{d}(s)^{\!\top} \boldsymbol{q}+ \boldsymbol{w}^k(s)^{\!\top}\boldsymbol{v}+\frac{\rho}{2}\!\parallel \! \boldsymbol{v} - \boldsymbol{\hat{v}}^k \! \parallel^2\}$
\renewcommand{\algorithmicwhile}{\textbf{while}}
\vspace{-0.2cm}
\EndWhile
\end{algorithmic}
\label{alg2}
\end{algorithm}

\section{Case Studies}
\subsection{Experiment Settings}
In this paper, we projected the 118-bus transmission systems in Texas, where ice storms occur frequently, for case studies. We modified those systems by incorporating wind farms. Specifically, four wind farms with a capacity of 500MW are added in bus 23, 70, 94 and 103. The historical data from the ERCOT market during the winter of 2020, a period that witnessed severe ice storms in Texas, is used for load and wind power scenario generation. To ensure comparability across cases, we fix the total simulation time $T$ in the second stage as 36 hours within a 1-hour resolution. The contingency operation duration $T_E$ is 24 hours and the preparation time $T_{N}$ varies from 2 hours to 12 hours respectively to analyze the impact of predictive information. The normal operation without knowing the occurrence of ice storms should be $T_0=T-T_E-T_{N}$. The objective function and operational constraints are the same as  (\ref{NormalPowerFlow})-(\ref{PtoEratioN}) except that no load shedding is allowed during this period. 

For investment parameters, the cost of line hardening is set as 1 million \$/mile. Generic community-scale energy storage is assumed and does not have specific restrictions on its placement in the transmission system. The capital costs of the energy storage are set to be 75\$/kWh and 1300\$/kW and are pro-rated on a daily basis using the net present value approach in \cite{ghofrani2013framework}. The lifetime of the energy storage is assumed to be 10 years with an annual discount rate of 5\%. The energy-to-power ratio is set to be 6 hours and both the charging and discharging efficiencies are set as 0.9 \cite{hassan2018energy}.


For operational parameters, the penalty costs of load shedding during contingencies and wind curtailment are assumed to be 2000\$/MWh and 500\$/MWh. 
The time-varying load-shedding penalty during preparation is assumed to be encapsulated using a monotonically decreasing exponential function: 
$c_{LS}^N(t) = a^{bt}+c$, where $a(a>1)$, $b$ and $c$ are parameters that indicate the confidence level in the forecasts. A higher penalty corresponds to greater uncertainty. 
The impact of varying levels of uncertainty in predictive information will be evaluated in the subsequent section.
The load-shedding costs of critical buses with higher priority are 2 times that of non-critical buses. For preventive load shedding, 10\% of the critical buses and 20\% of normal buses are allowed. During ice storms, 20\% critical load shedding is allowed and the normal load shedding will be flexible since the operational objective will be changed to retain the critical load. The procedure in section III is employed to generate contingency scenarios. 
The probabilities and intensities of ice storms can be estimated using historical data.

All models and algorithms are implemented using Julia. All the stochastic mixed-integer programming problems are solved using the Gurobi solver on a PC running the 64-bit Windows operating system with AMD Ryzen 9 5900X CPU clocking at 3.7 GHz and 64 GB of RAM. The relative convergence gap has been set to 1\%. 

\subsection{Performance of the Two-stage RIDDRP model}
To evaluate the efficiency of the proposed RIDDRP model, which incorporates the information and DDU, we compared the amount of load shedding during ice storms of four resilience enhancement strategies with different considerations, as shown in Table \ref{CompareExper}. Strategy I is the proposed model. Strategy II sets the preparation time $T^N$ as 0, implying no consideration of predictive information. Strategy III adopts a constant preventive load-shedding penalty, disregarding the uncertainty in predictive information. Strategy IV ignores the DDU and excludes line-hardening strategies. The results are shown in Table \ref{CompareLSCost}.
\begin{table}[]
\small
\centering
\caption{Planning strategies with different considerations} 
\vspace{-0.5em}
\renewcommand{\arraystretch}{1.1}
\begin{tabular}{cccc}
\hline \hline
Strategy & \begin{tabular}[c]{@{}c@{}}Anticipatory\\ Preparation\end{tabular} & \begin{tabular}[c]{@{}c@{}}Prediction\\ Uncertainty\end{tabular} & \begin{tabular}[c]{@{}c@{}}Decision-\\ Dependency\end{tabular} \\ \hline
I        & $\checkmark$                                                                   & $\checkmark$                                                                & $\checkmark$                                                              \\
II       & $\times$                                                                  & $\setminus$                                                 & $\checkmark$                                                              \\
III      & $\checkmark$                                                                  & $\times$                                                                & $\checkmark$                                                              \\
IV       & $\checkmark$                                                                  & $\checkmark$                                                                & $\times$                                                              \\ \hline \hline
\end{tabular}
\label{CompareExper}
\vspace{-0.5em}
\end{table}

\begin{table}[t]
\small
\centering
\caption{Load-shedding costs of different strategies (\$)} 
\vspace{-0.5em}
\renewcommand{\arraystretch}{1.1}
\begin{tabular}{ccc}
\hline \hline
Strategy & Preparation       & Emergency        \\ \hline
I        & 149146.12      & 8864831.45       \\
II       & $\setminus$ & $\setminus$ \\
III      & 203537.94       & 8808182.29       \\
IV       & 193033.51       & 9126962.12       \\ \hline \hline
\end{tabular}
\label{CompareLSCost}
\vspace{-1em}
\end{table} 

Under the given storage budget, strategy II fails to meet critical load-shedding limits, resulting in infeasible solutions. This underscores the importance of considering predictive information for prior preparation. Strategy III ignores the time-varying forecasting uncertainty, leads to a substantial increase in preventive load-shedding. Notably, this increase in cost is significantly higher than the reduction in emergency load-shedding, rendering it an inferior solution. Strategy IV ignores the benefits of line hardening and yields the highest costs in both preventive and emergency load-shedding. This indicates the necessity of considering the effect of robust system design to further enhance system resilience. 
 

\subsection{Risk-informed Resilience Enhancement Outcomes}
\begin{figure}[b]
\vspace{-0.7em}
\centering
\includegraphics[scale=0.35]{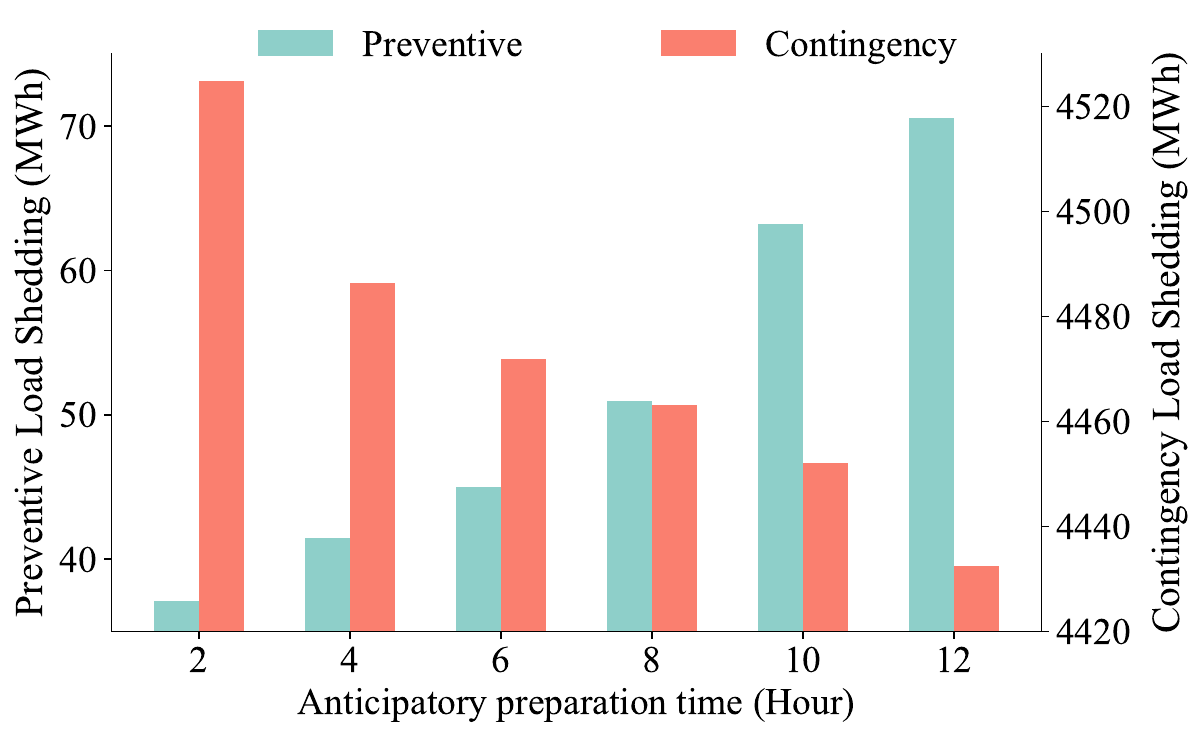}
\vspace{-0.5em}
\caption{Preventive and contingency load shedding in IEEE-118 bus system.}
\label{LS118}
\end{figure}
As mentioned in Section II, predictive information can influence risk-informed operations by influencing both the time and dispatchable resources available to operators. In this study, we investigate the resilience of the test system under each investment plan considering diverse uncertainty levels of predictive information, which is parameterized by the preparation time and the preventive load-shedding penalty. Fig. \ref{LS118} illustrates the load shedding during both the risk-informed preparation phase and ice storms in the test system. Notably, as the preparation time increases, the preventive load shedding increases while the load shedding during ice storms decreases. For instance, elevating the preventive non-critical load shedding by 33.4 MWh can lead to a substantial reduction of 92.4 MWh in load shedding amidst ice storms. This demonstrates that even a marginal amount of preventive load shedding can precipitate a multiple times decrease in load shedding during contingencies, as observed in the case studies, and potentially even higher for larger systems, underscoring the imperative of incorporating risk-informed operations prior to anticipated events.

\begin{figure}[!t]
    \centering
    \includegraphics[scale=0.37]{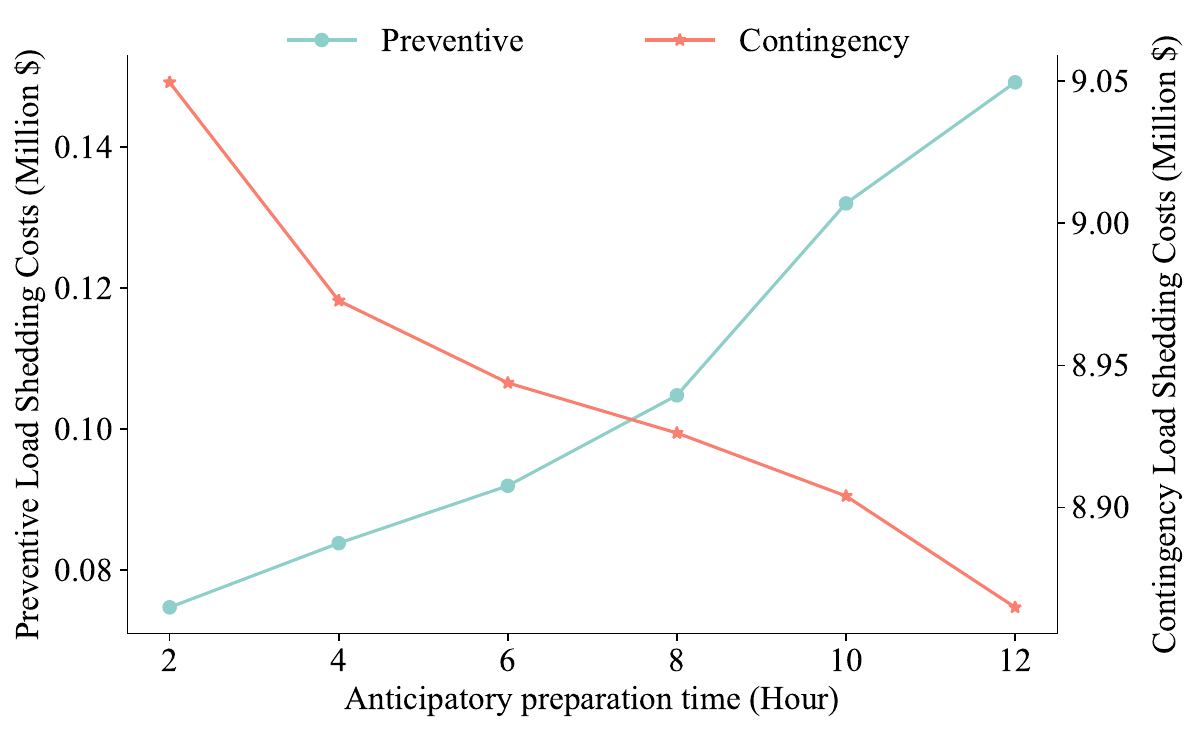}
    \vspace{-0.5em}
    \caption{Load shedding costs in IEEE-118 bus system.}
    \label{Cost118}
    \vspace{-0.5em}
\end{figure}

Given that we map the uncertainty of predictive information into time-varying preventive load-shedding costs, the expected preventive and contingency load-shedding costs under various levels of predictive information uncertainty are presented in Fig. \ref{Cost118}. Despite incurring a higher load-shedding penalty prior to anticipated ice storms, the benefits of preventive load-shedding are discernible. A relatively modest increase in advance load-shedding costs can precipitate a more substantial reduction in costs when navigating higher demand during ice storms. Thus, the incorporation of predictive information can enable a more resilient operation in the face of extreme events.

\begin{figure}[!t]
    \centering 
    \vspace{-0.5em}  
    \includegraphics[scale=0.45]{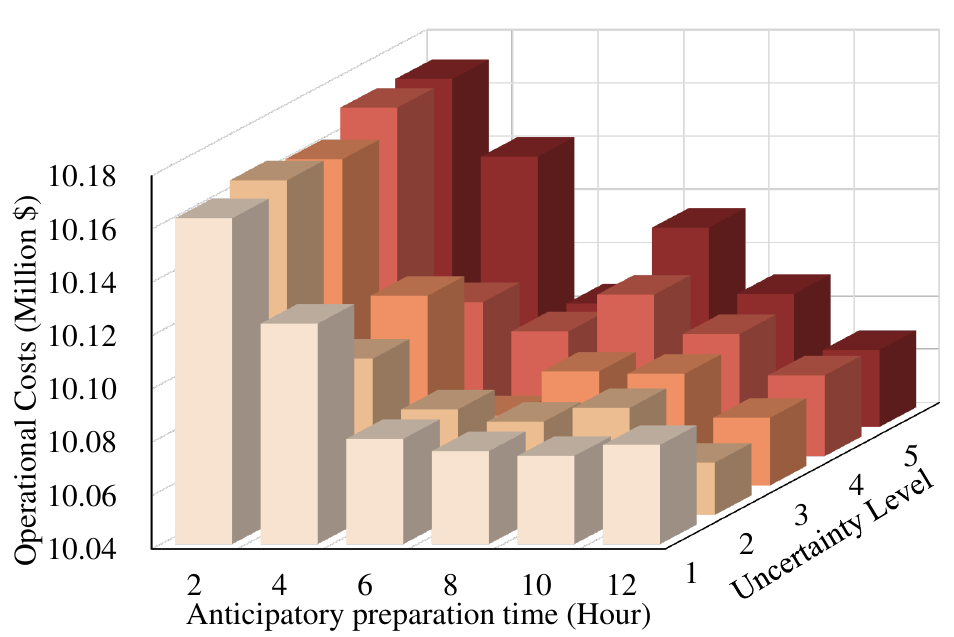}
    \vspace{-0.5em}
    \caption{Operational costs under various preparation time and forecasting confidence levels in IEEE 118 system.}
    \label{Uncertainty}
    \vspace{-0.5em}
\end{figure}

Furthermore, we investigate the impact of the confidence level in the predictive information on resilience planning. We selected five different load-shedding penalty curves, each represented by an exponential function with varying decay parameters, to simulate different levels of time-varying forecasting accuracy. These curves all converge to a penalty of 2000\$/MWh, which is the designated load-shedding penalty during ice storms. A faster decay implies lower forecasting accuracy and higher uncertainty,
thereby increasing the risks associated with performing preventive load-shedding. The results are shown in Fig. \ref{Uncertainty}. Similarly, operational costs decrease as the preparation time increases across all uncertainty levels. However, under a fixed preparation time, the increase in operational costs due to changes in the confidence level is less pronounced. This indicates that the proposed model is less sensitive to the variances in the time-varying confidence level in this case.



\vspace{-1em}
\subsection{The Value of Predictive Information}
\begin{figure}[!t]
\centering
\includegraphics[scale=0.4]{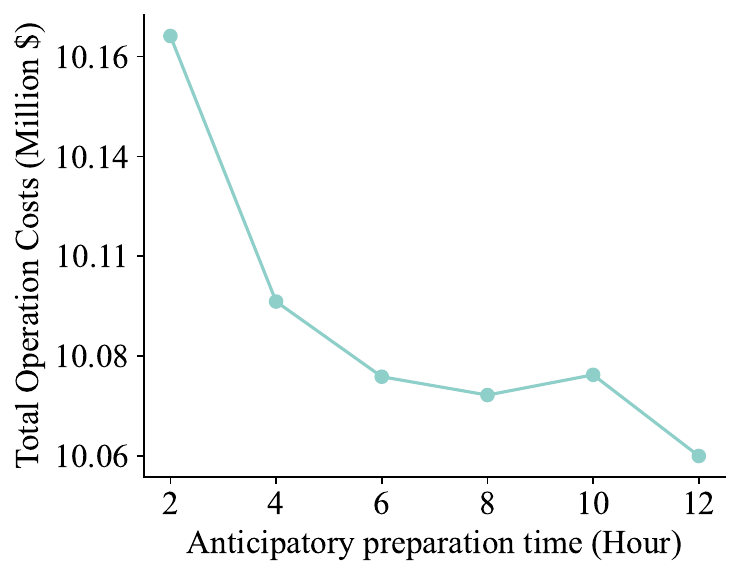}
\vspace{-0.5em}
\caption{Total operational costs under various anticipatory preparation time.}
\label{OperationTotal}
\vspace{-0.5em}
\end{figure}

As mentioned above, the strategic utilization of predictive information presents a balance between enhancing system resilience and incurring additional costs due to the time-varying uncertainty inherent in predictive information. 
To quantify the value of predictive information, in this section, we first investigate the total operation costs during the entire operational horizon, which maintains consistency for all cases. The results are shown in Fig. \ref{OperationTotal}. It can be noticed that even though the uncertainty level of predictive information may increase as the time until the anticipated ice storms lengthens, the total operational costs can still witness a reduction of \$105,951 by transitioning to risk-informed operation by taking preventive actions more proactively. 
As the anticipatory preparation time lengthens, initiating preventive load-shedding earlier incurs higher penalty costs. Nonetheless, when juxtaposed with the potential cost benefits during contingencies, accepting slightly elevated risks to proactively make preparations can pave the way for enhanced system resilience in the face of ice storms.

 
Then, we calculate the marginal cost increments of preventive load shedding associated with extending the anticipatory preparation time in the IEEE-118 system, which is shown in Table \ref{MarginalCost}.
It can be noticed that there is a discernible increase in marginal costs with extended preparation times. For instance, the marginal cost for extending preparation from 6 to 8 hours is \$12,824.61, while it more than doubles to \$27,207.76 for an extension from 8 to 10 hours. 
The results reveal a critical point where the cost of additional preparation outweighs the benefits. While longer preparation enables more comprehensive measures to mitigate load shedding during emergencies, the associated higher marginal costs underscore the need for a judicious selection of preparation time, balancing the benefits of enhanced resilience against escalating costs. 

\begin{table}[]
\small
\centering
\caption{Marginal preventive operational cost increment} 
\renewcommand{\arraystretch}{1.1}
\begin{tabular}{cc}
\hline \hline
\begin{tabular}[c]{@{}c@{}}Anticipatory Preparation\\Time (Hour)\end{tabular} & \begin{tabular}[c]{@{}c@{}}Marginal Costs (\$)\end{tabular} \\ \hline  
2                                   & $-$                             \\
4                                   & 9093.14                      \\
6                                   & 8140.09                      \\
8                                   & 12824.61                      \\
10                                   & 27207.76                      \\
12                                   & 17169.74                      \\ \hline \hline
\end{tabular} 
\label{MarginalCost}
\vspace{-1em}
\end{table}

\subsection{Balancing Investment Costs and System Resilience via Risk-informed Planning}
In this section, we investigate the relationship between investment decisions and risk-informed operations influenced by predictive information. 
For line-hardening decisions, as the anticipatory preparation time extends 
from 2 hours to 12 hours, the hardening costs are 2901.96, 2953.91, 2836.15, 2953.91, 2865.97, 2865.97 Milion \$ respectively.
Initially, the system planner is equipped with enhanced insights into system operation, potentially necessitating more line hardening to preemptively mitigate load shedding.
This can be observed from the results that the line hardening costs increase from 2901.96 million \$ to 2987.93 million \$ as the anticipatory preparation time extends from 2 to 6 hours.
However, as this time extends to 12 hours, the line hardening costs decrease and converge to 2865.97 million \$. This can be explained by the fact that there are more dispatchable resources available, i.e., energy storage in this study, thereby enhancing the system's flexibility in managing anticipated contingencies by achieving a trade-off between uncertain information and potential disruptions. Under the circumstances, the system can now leverage more economically viable alternatives to alleviate the impacts of potential system failures. Thereby, the imperative to harden additional lines diminishes. 
Based on the results, this dynamic interplay between enhanced predictive information, resource availability, and system flexibility underscores the intricate balance that system planners and operators must navigate to optimize investment costs while ensuring resilient system operation in the face of anticipated disruptive events.


\begin{figure}[!t]
\centering
\vspace{-0.8em}
\includegraphics[scale=0.4]{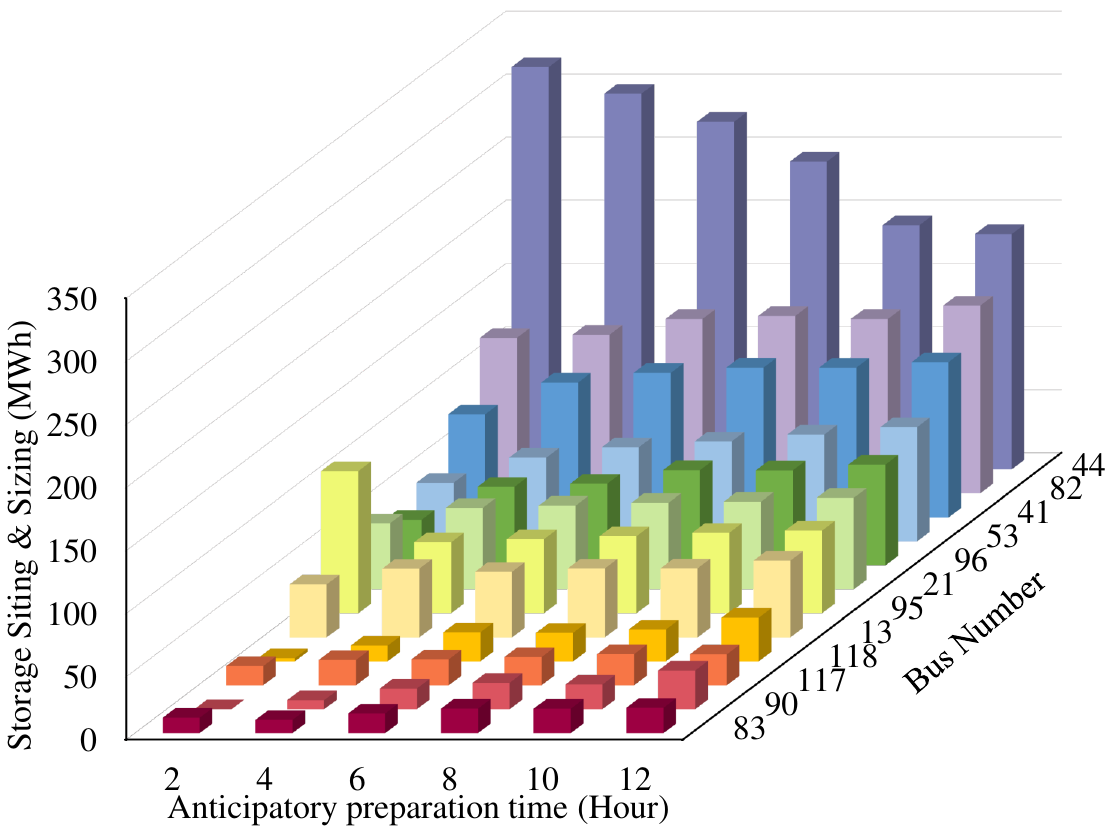}
\vspace{-0.3em}
\caption{Illustrative example of siting and sizing results of energy storage in IEEE 118-bus System.}
\vspace{-1em}
\label{Storage118}
\end{figure}

The impact of varying predictive information on energy storage planning decisions is illustrated in Fig. \ref{Storage118}, which presents some energy storage siting and sizing results in the IEEE 118-bus system. While the siting decisions are almost the same among different cases, the allocation varies significantly given different anticipatory time. 
For instance, for Bus 96, the sizing of the energy storage evolves from about 36 MWh to 80 Mwh, revealing a significant increase. Conversely, Bus 95 shows a different pattern, with sizing decisions decreasing from about 112 MWh to 65 MWh as risk-informed operation stages increase from 2 to 12 hours. This indicates that by taking into account the influence of the predictive information, the system planner can achieve a more risk-informed allocation of resources, resulting in a synergistic balance between economic and resilient operations in the face of future extreme events.

\section{Conclusion}
Recognizing the pivotal role of predictive information, this paper proposed a two-stage risk-informed decision-dependent resilience planning (RIDDRP) model against ice storms. The RIDDRP model effectively integrates strategic line hardening and energy storage planning decisions to achieve risk-informed operation considering both normal and emergency conditions. The model captures the DDU of contingency distributions introduced by line hardening decisions, the inherent uncertainty in predictive information, and the exogenous uncertainties associated with ice storms. To alleviate the computational burdens, PHA is employed to solve large-scale mixed-integer programming problems. 
Numerical studies reveal that the RIDDRP model can enhance system resilience from two aspects. Firstly, it informs investment decisions by facilitating a more judicious allocation of resources considering multi-source uncertainties. Secondly, it leverages the predictive information to facilitate risk-informed coordination between normal operations, which can make proactive preparations, and emergency operations during contingencies, by optimizing the utilization of dispatchable resources. 
Furthermore, the value of predictive information is quantified to analyze its impacts on planning and operation. This provides insights into identifying more influential properties for risk-informed operations. 
A key advantage of the proposed model is the incorporation of predictive information at the planning stage, which enables system planners to make resilient investment decision-making through a more efficient allocation of resources.
Consequently, this approach fosters a harmonious balance between economic efficiency and resilience, equipping power systems to better withstand and recover from future extreme events.

\bibliographystyle{ieeetr}
\bibliography{ref}


\vfill

\end{document}